\documentclass[conference]{IEEEtran}
\IEEEoverridecommandlockouts

\usepackage{subcaption}
\usepackage{graphicx}
\usepackage{amsmath,amssymb,amsfonts}
\usepackage{bm}
\usepackage{color, soul}
\usepackage{multirow}
\usepackage{soul}
\usepackage{balance}
\usepackage{cite}
\usepackage{url}

\newcommand{\smartpara}[1]{\noindent \textbf{#1.}}

\usepackage{titlesec}
\titlespacing\section{0pt}{5pt}{2.5pt}
\titlespacing\subsection{0pt}{3pt}{2pt}
\def\BibTeX{{\rm B\kern-.05em{\sc i\kern-.025em b}\kern-.08em
    T\kern-.1667em\lower.7ex\hbox{E}\kern-.125emX}}

\begin{document}

\title{[Experiment, Analysis, and Benchmark] Systematic Evaluation of Plan-based Adaptive Query Processing}

\author{
\IEEEauthorblockN{Pei Mu}
\IEEEauthorblockA{\textit{University of Edinburgh} \\
United Kingdom \\
pei.mu@ed.ac.uk \\
0000-0003-2781-8032}
\and
\IEEEauthorblockN{Anderson Chaves Carniel}
\IEEEauthorblockA{\textit{Huawei Technologies Research \&} \\ \textit{Development (UK) Limited} \\
United Kingdom \\
anderson.chaves.carniel@huawei.com \\
0000-0002-8297-9894}
\and
\IEEEauthorblockN{Antonio Barbalace}
\IEEEauthorblockA{\textit{University of Edinburgh} \\
United Kingdom \\
antonio.barbalace@ed.ac.uk \\
0000-0003-1641-0779}
\and
\IEEEauthorblockN{Amir Shaikhha}
\IEEEauthorblockA{\textit{University of Edinburgh} \\
United Kingdom \\
amir.shaikhha@ed.ac.uk \\
0000-0002-9062-759X}
}








\maketitle

\begin{abstract}
Unreliable cardinality estimation remains a critical performance bottleneck in database management systems (DBMSs). Adaptive Query Processing (AQP) strategies address this limitation by providing a more robust query execution mechanism. Specifically, plan-based AQP achieves this by incrementally refining cardinality using feedback from the execution of sub-plans. However, the actual reason behind the improvements of plan-based AQP, especially across different storage architectures (on-disk vs. in-memory DBMSs), remains unexplored.

This paper presents the first comprehensive analysis of state-of-the-art plan-based AQP. We implement and evaluate this strategy on both on-disk and in-memory DBMSs across two benchmarks. Our key findings reveal that while plan-based AQP provides overall speedups in both environments, the sources of improvement differ significantly. In the on-disk DBMS, PostgreSQL, performance gains primarily come from the query plan reorderings, but not the cardinality updating mechanism; in fact, updating cardinalities introduces measurable overhead. Conversely, in the in-memory DBMS, DuckDB, cardinality refinement drives significant performance improvements for most queries. We also observe significant performance benefits of the plan-based AQP compared to a state-of-the-art related-based AQP method.
These observations provide crucial insights for researchers on when and why plan-based AQP is effective, and ultimately guide database system developers on the tradeoffs between the implementation effort and performance improvements.

\end{abstract}

\section{Introduction}\label{section:introduction}

The increasing prevalence of data characterized by significant diversity in type and distribution necessitates efficient processing techniques. In real-world industry scenarios, companies often need to integrate or analyze complex data using online analytical processing (OLAP) queries, which involve heavy and sophisticated joins. Database management systems (DBMSs) are used to handle complex OLAP workloads. The traditional architecture of DBMSs uses a query optimizer to select the optimal execution plan, followed by a query execution engine. 
The query optimizer enumerates join order candidates and chooses the most efficient alternative using the estimated cardinality as the primary input to its cost model.

The traditional architecture of DBMS can always take the best execution plan as long as the cardinality estimation and the cost model are accurate. However, as the number of joins increases, DBMSs typically produce large estimation errors, which is often the reason for suboptimal plans~\cite{leis2015good}. For example, DuckDB (v0.10.1)~\cite{duckdb_code} executes one of the Join Order Benchmark (JOB)~\cite{leis2015good} queries (\texttt{8d.sql} with $8$ joins) in \boldsymbol{$2.7$} seconds. However, with accurate cardinalities this query can be executed in only \boldsymbol{$0.8$} seconds; DuckDB achieves a \boldsymbol{$3.3 \times$} speedup by just correcting the cardinality estimation.

Over the last three decades, a spectrum of adaptive query processing (AQP) methods~\cite{babu2005adaptive, deshpande2007adaptive, deshpande2006adaptive} has been proposed to address the challenges of inaccurate cardinality estimation. 
A class of such techniques refine cardinality estimation via the execution of small tuple sets or random samples~\cite{bizarro2006adaptive, li2006adaptively, ives2004adapting, babu2005proactive, zhu2004dynamic, babu2004adaptive}. However, they face substantial overhead due to query plan monitoring, which becomes prohibitive for queries with many joins.
There have also been recent developments based on reinforcement learning methods~\cite{kaftan2018cuttlefish, trummer2021skinnerdb}. However, such techniques suffer from the overhead of the heavy learning process, and there is a risk of overfitting. 

Plan-based AQP methods~\cite{kabra1998efficient, markl2004robust, bruno2002exploiting, neumann2013taking, karanasos2014dynamically, zhao2023efficient, herlihy2024adaptive} are more lightweight and robust for queries with multiple joins. These AQP methods divide the initial query into sub-plans and while running sub-plans, they update the cardinality of the next sub-plans. This allows the DBMS to use more accurate cardinalities for the next sub-plans. Moreover, plan-based AQP usually does not require modifications to the core modules of the DBMS, such as the optimizer and executor. This makes modular expansion and application to existing DBMSs easier. 

Although the recent research shows performance improvements from the plan-based AQP strategy, there is no clear indication of the contribution of different components.
Despite evidence that slight cardinality estimation errors can sometimes lead to substantial slowdowns in execution time, the performance effect (either positive or negative) of refining the cardinality is still an open question~\cite{bergmann2025elephant}.

This paper aims to analyze plan-based AQP systematically. We aim to understand the performance benefits of each component of plan-based AQP. Based on this, we can provide guidelines for database system developers on the trade-off between the implementation effort required for each component and the performance benefits it can offer.

We adopt QuerySplit~\cite{zhao2023efficient} as a representative state-of-the-art (SOTA) and benchmark it through a comprehensive module-based performance decomposition and analysis. 
QuerySplit combines the heuristic join reordering rules and the refinement of the cardinalities adaptively during the execution. As a result, the source of the performance improvement for QuerySplit remains unclear.
We observed that for most queries from the JOB and Decision Support Benchmarks (DSB)~\cite{ding2021dsb}, the speedup from QuerySplit is achieved not by updating the cardinality, but rather through join reordering, which is an optimization orthogonal to plan-based AQP.

While this paper focuses on the analysis of plan-based AQP, we want to understand whether it has performance advantages compared with other AQP methods. We compared the performance of the SOTA plan-based AQP with a SOTA relation-based AQP, Polar~\cite{justen2024polar}, which updates cardinality from the execution of subsets of relations. We observed that for most queries from the JOB benchmark, the plan-based AQP achieves a better performance.

These observations lead us to ask the following research questions: How effective is the SOTA plan-based AQP in comparison with the traditional architecture? How is its applicability to the in-memory DBMS? What impact does each component have on performance? And how good is the plan-based AQP compared to the SOTA relation-based AQP?

\smartpara{Core Contributions} To answer these questions, we present a systematic study of the SOTA plan-based AQP, including analyzing the performance breakdown by isolating different components. 
We consider QuerySplit, which is based on PostgreSQL and is an on-disk DBMS.
In addition, we implement plan-based AQP on top of an in-memory DBMS, DuckDB, with a new modular design.
This modular design gives us the flexibility to isolate different components and facilitates the transplantation of the plan-based AQP idea to other open-source systems. As a result, database practitioners can decide which AQP strategy is more feasible and easier to implement for their existing database engines. For evaluation purposes, we measured performance on different benchmarks and at various scale factors to demonstrate generality and scalability, and implemented the plan-based AQP on three different versions of DuckDB (v0.6.1, v0.10.1, and v1.3.2) to show the flexibility.

\smartpara{Summary of Results}
Based on our experimental results for the on-disk DBMS, the main speedup comes from plan reordering, rather than correcting the cardinality. However, for the in-memory DBMS, the updated cardinalities help the optimizer to come up with significantly improved sub-plans. 
Although we observe that SOTA plan-based AQP methods still have shortcomings and opportunities for further performance improvement, plan-based AQP performs better than the relation-based AQP, especially for queries with complex joins.

\smartpara{Detailed Contributions and Outline} 
After giving the background and reviewing the related work in Section~\ref{section:preliminaries}, we make the following contributions:
\begin{itemize}
  \item We provide a unified representation to guide database system developers on the trade-off between the implementation effort and performance gains in Section~\ref{design:components}.
  \item We propose a modular design and module-level isolation extension for plan-based AQP with an on-disk DBMS in Section~\ref{design:postgres} and an in-memory DBMS in Section~\ref{design:duckdb}.
  \item We study the performance impact of each plan-based AQP module of the two DBMS systems using the setups in Section~\ref{section:setup} and experimental evaluations in Section~\ref{section:eval}\footnote{https://github.com/PeiMu/Reproducibility-Evaluating-Plan-based-Adaptive-Query-Processing}.
\end{itemize}

Finally, we summarize the key takeaways in Section~\ref{section:conclusion}.



\section{Preliminaries and Related Work}\label{section:preliminaries}

This section outlines the overall foundations for our paper by (i) describing the problem of suboptimal plans generated by the traditional DBMSs, both on-disk and in-memory, due to the inaccurate cardinality estimation (Section~\ref{preliminaries:on_disk_and_in_memory}), and (ii) reviewing the adaptive query processing (AQP) methods that aims at solving this problem (Section~\ref{preliminaries:aqp}).

\subsection{Cardinality Estimation Issue for DBMSs}\label{preliminaries:on_disk_and_in_memory}
The inaccurate cardinality estimation is a known issue for traditional DBMSs~\cite{lan2021survey}, where the optimizer and the cost model work together to find the most efficient execution plan based on the estimated cardinality as input. The cost model estimates the cost of each operator within the physical plan, while the optimizer selects the physical plan with the lowest total estimated cost. However, as pointed out by Leis et al.~\cite{leis2015good}, cardinality estimation is often inaccurate, and minor errors are propagated exponentially through joins, which can lead to suboptimal and sometimes disastrous plans.

While the core challenge of inaccurate cardinality estimation affects both on-disk and in-memory DBMSs, its performance impact profile differs significantly due to the distinct bottlenecks of each. In on-disk DBMSs, the dominant cost is often I/O. Suboptimal plans induced by cardinality errors can result in excessive disk accesses or inefficient buffer utilization, leading to a drastic increase in execution time. Additionally, due to the high I/O cost, on-disk DBMSs typically incur significant overhead when creating a temporary table to store intermediate results and reading the table when reusing it.
In contrast, in-memory DBMSs eliminate the disk I/O bottleneck by storing the metadata and intermediate results in memory, but shift the primary cost to CPU processing and memory access. Incorrectly selected plan orders can generate unexpectedly large intermediate results, resulting in inefficient CPU utilization and the risk of memory spills. 
PostgreSQL~\cite{stonebraker1986design} and DuckDB~\cite{raasveldt2019duckdb} are representative database engines of on-disk and in-memory DBMSs, respectively. To investigate the performance influence, our experimental study in Section~\ref{section:eval} is measured on top of PostgreSQL and DuckDB.

\subsection{Adaptive Query Processing}\label{preliminaries:aqp}
To address the suboptimal performance issue of inaccurate cardinality estimation, various AQP strategies are proposed to collect ``actual statistics'' during execution and optimize the plan based on the updated statistics. We present a survey of the existing AQP-related work in Section~\ref{preliminaries:survey} and describe the SOTA in Section~\ref{preliminaries:query_split}. 

\subsubsection{Survey of Adaptive Query Processing}\label{preliminaries:survey}

AQP has been explored in the past few decades. The existing  surveys~\cite{babu2005adaptive, deshpande2007adaptive, deshpande2006adaptive, haritsa2024robust} classify the AQP strategy into two broad categories: \textit{inter-query AQP}, which uses observation or knowledge learned from previous query executions to guide the estimation of the cost of the subsequent queries; and \textit{intra-query AQP}, which reacts to query workload, cost, or data changes immediately in the current query plan. 

The intra-query AQP can be classified into two sub-categories: relation-based intra-query AQP, which partitions the relation into chunks and gradually refines the cardinality after executions of several partitions; and plan-based intra-query AQP, which splits the plan into several sub-plans and increasingly optimizes each sub-plan with the ``actual cardinality'' gained from the execution of the last sub-plan.

\smartpara{Inter-query AQP} Adaptive datalog~\cite{chen1994adaptive, aboulnaga1999self, bruno2001stholes} and adaptive monitor~\cite{stillger2001leo, haas2005automated, bizarro2008progressive} fall into this category. ASE~\cite{chen1994adaptive} uses a curve-fitting function to correct the real attribute value distribution. ST-histograms~\cite{aboulnaga1999self} and STHoles~\cite{bruno2001stholes} construct custom histograms and update them when a significant estimation error occurs. LEO~\cite{stillger2001leo} adds a new component, \textbf{monitor}, inside the optimizer to check the cardinality of each operator in the query plan. ASC~\cite{haas2005automated} monitors the compile-time information of a query (plan monitor), runtime information about the statistics (runtime monitor), and the number of rows for a table (activity monitor). PQO~\cite{bizarro2008progressive} constructs a special data structure (called a parametric plan) to maintain the optimal physical plans and monitor their cardinality and cost. When a new query arrives, it decides whether the history plans can be reused or need to generate a new physical plan.

\smartpara{Relation-based Intra-query AQP} For the relation-based AQP, tuple routing is one of the most critical strategies, involving the continuous reordering of operators in a query plan during its execution. Eddies~\cite{avnur2000eddies} is the first work of this kind, which introduces a new operator to distribute tuples and reorder operators for different tuples on the fly continuously. SteM~\cite{raman2003using}, SHARP~\cite{bizarro2006adaptive}, ARJ~\cite{li2006adaptively}, HPE~\cite{tzoumas2010sharing}, CACQ~\cite{madden2002continuously}, and STAIRs~\cite{deshpande2004lifting} extend this work by incorporating a tuple router, making the system more applicable. Specifically, SHARP and ARJ also have a monitor module to collect statistics from the intermediate operators. This idea has also been applied to distributed settings~\cite{tian2003tuple}. 

There are other strategies of the relation-based AQP method. ADP~\cite{ives2004adapting} partitions data adaptively and refines the plan for different data regions. Rio~\cite{babu2005proactive} generates robust and switchable plans based on the bounding boxes and re-optimizes the plan if the actual cardinality exceeds the bound. A-Greedy~\cite{babu2004adaptive} continuously monitors the cardinality of the intermediate operators using a random sample over the recent past. CAPE~\cite{zhu2004dynamic} follows a similar process, focusing on the plan-changing step by mapping and moving all relevant tuples from the old plan to the new plan.
Cuttlefish~\cite{kaftan2018cuttlefish} and SkinnerDB~\cite{trummer2021skinnerdb} use reinforcement learning methods to tune the join orders.

\smartpara{Plan-based Intra-query AQP} The main idea behind plan-based AQP is to find the appropriate point to revisit the cardinality of the plan while executing it. Dynamic Reoptimization~\cite{kabra1998efficient} and POP~\cite{markl2004robust} set one or more checkpoints in the query plan and decide if re-optimization is needed by checking the statistics. SIT~\cite{bruno2002exploiting} statically decides which key intermediate operators are important and adaptively updates the cardinality of these specific operators only to reduce the extra cost. IEF~\cite{neumann2013taking} halts the query execution at the pre-determined point in the plan and updates the statistics to remove the uncertainty, then re-optimizes it based on the actual statistics. DYNO~\cite{karanasos2014dynamically} splits the plan at the materialization point and executes the sub-plan over a sample of data to refine the statistics. Carac~\cite{herlihy2024adaptive} focuses on the recursive query execution and considers different plans for each recursive iteration. There are also efforts to execute sub-plans in parallel~\cite{condon2006flow}; however, these efforts are limited by the initial sub-plan, especially when the cardinality is often unknown and inaccurate at the outset. 

\subsubsection{SOTA of Plan-based Adaptive Query Processing}\label{preliminaries:query_split}

Plan-based AQP demonstrates particular efficiency for multi-join queries, where cardinality estimation errors most severely degrade performance. This is mainly thanks to its theoretical capacity for progressive refinement. Furthermore, plan-based AQP offers practical implementation advantages with non-intrusive integration requiring only targeted extensions to existing engine components. 

QuerySplit~\cite{zhao2023efficient} is the SOTA for plan-based AQP. This system avoids the possible defects of the initial plan that are described in Section~\ref{preliminaries:survey} and decreases the damage of the inaccurately estimated cardinality by splitting the plan into sub-plans and executing the smallest sub-plan first and postponing the larger sub-plans. 

\begin{figure}[t]
    \centering
    \includegraphics[width=.7\columnwidth]{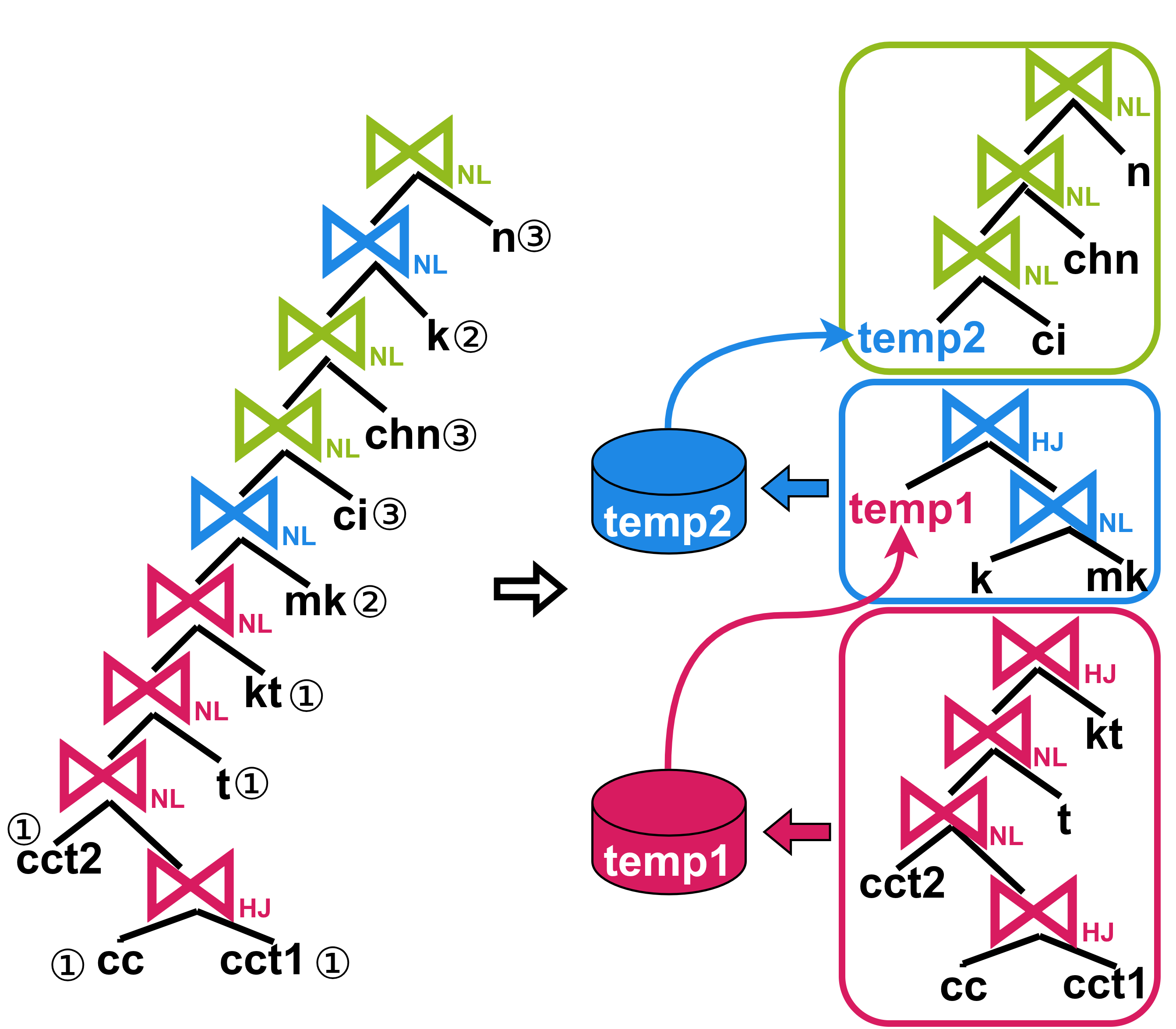}
    \caption{An example of the QuerySplit, where the left part is the logical plan of \texttt{20c.sql} from the JOB benchmark, the right part is the three sub-plans generated. The numbers beside the relations indicate which sub-plan it belongs to and also represent the order in which the sub-plan is executed.}
    \label{fig:2_example}
\end{figure}

\begin{figure}[t]
    \centering
    \includegraphics[width=.55\columnwidth]{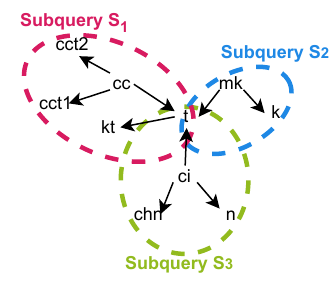}
    \vspace{-0.1cm}
    \caption{DAG generated by the plan-based AQP method of \texttt{20c.sql} from the JOB benchmark. It has three sub-plans, where the split point is the relation \texttt{t}.}
    \label{fig:2_example_dag}
\end{figure}

\smartpara{Mechanism} At a high level, the QuerySplit system splits and selects the sub-plans on top of the logical plan represented by a directed acyclic graph (DAG).
An example (cf. Figure \ref{fig:2_example}) of the \texttt{20c.sql} from the JOB benchmark explains how the QuerySplit works. Firstly, the QuerySplit system constructs a DAG (cf. Figure \ref{fig:2_example_dag}), where the vertices represent the relations and the edges represent the join predicates. The arrow points from the relation with the foreign key to the relation with the primary key.
When there is a relation pointed by more than one arrow (e.g., relation \texttt{t} in Figure~\ref{fig:2_example_dag}), this relation is considered as the split point. 
QuerySplit executes the sub-plan with the lowest cost and then generates a temporary table to store the intermediate result. It is worth noting that this step is based on the heuristics join ordering rules, which prioritize the smaller joins.
Next, QuerySplit collects and updates the ``actual'' cardinality of each sub-plan by actively calling the \texttt{ANALYZE} process in Postgres, and will use it to help with the optimization of the following sub-plans.

\smartpara{Limitations} The QuerySplit method has several limitations, by design. Firstly, the DAG representation naturally requires foreign key information to construct the graph's direction and cannot function without foreign key constraints. 
Second, its splitting strategy requires that the splitting point is the relation with the primary key (e.g., relation \texttt{t} in Figure~\ref{fig:2_example_dag}), where the DAG can construct multiple sub-plans centered on the relation containing the foreign key (e.g., relations \texttt{cc}, \texttt{mk}, and \texttt{ci} in Figure~\ref{fig:2_example_dag}). Thus, QuerySplit cannot split plans for queries with a star(-like) schema. Thirdly, the DAG representation loses non-SPJ operators and cannot perform well for queries that involve such operators.
Lastly, the graph-based plan representation requires traversing every possible sub-graph (sub-plans) and suffers from the performance overhead.

\subsubsection{SOTA of Relation-based Adaptive Query Processing}\label{preliminaries:polar}
Relation-based AQP shows performance improimprovements when the dataset is large and complex (e.g., skewed). It achieves adaptivity by splitting the relation rather than the plan. As a result, relation-based AQP can achieve considerable speedups when the query is simple and involves few joins.

Polar~\cite{justen2024polar} is the SOTA for relation-based AQP. This system introduces a self-regulating selection of alternative join orders and provides bounds for the overhead of adaptivity. Compared to the Eddies-inspired efforts that are described in Section~\ref{preliminaries:survey}, Polar is non-invasive to the optimizer and runtime, and it does not require state management. This dramatically simplifies the integration into existing DBMSs.

\smartpara{Mechanism} At a high level, the Polar system splits the relation into sets of tuples. During runtime, it routes different sets of tuples with each join order alternative, and decides the most efficient path based on performance indicators, e.g., the number of intermediates. Once the best join order is selected, the remaining part of the relation will be run. Furthermore, it processes a pipeline on each thread with an isolated state and shares build sides to probe, which improves scalability without incurring excessive additional performance overhead.

\smartpara{Limitations} Polar has several limitations. First, it only supports left-deep plans but not bushy plans. Secondly, Polar does not support non-SPJ operators but only supports pipelines with join sequences. Last but not least, it lacks the ability to change the source table when selecting candidate pipelines, which limits opportunities for better performance. 



\section{Module-level Design of Plan-based AQP}\label{section:module_level_design}

\begin{table*}[!htbp]
\caption{Unified module-level summarization of AQP methods.}\label{tab:unified_summary}
\begin{footnotesize}
\begin{tabular}{|l|l|l|ccccc|}
\hline
\multicolumn{1}{|c|}{\multirow{2}{*}{\textbf{Category}}} &
  \multicolumn{1}{c|}{\multirow{2}{*}{\textbf{Idea}}} &
  \multicolumn{1}{c|}{\multirow{2}{*}{\textbf{Method}}} &
  \multicolumn{5}{c|}{\textbf{Modification Required in Modules}} \\ \cline{4-8} 
\multicolumn{1}{|c|}{} &
  \multicolumn{1}{c|}{} &
  \multicolumn{1}{c|}{} &
  \multicolumn{1}{c|}{\textbf{Catalog}} &
  \multicolumn{1}{c|}{\textbf{Monitor}} &
  \multicolumn{1}{c|}{\textbf{Tuple Router}} &
  \multicolumn{1}{c|}{\textbf{Optimizer}} &
  \textbf{Plan Splitter} \\ \hline
\multirow{2}{*}{inter query} &
  \begin{tabular}[c]{@{}l@{}}adaptive\\ histogram\end{tabular} &
  \begin{tabular}[c]{@{}l@{}}ASE~\cite{chen1994adaptive}, ST-histograms~\cite{aboulnaga1999self},\\ STHoles~\cite{bruno2001stholes}\end{tabular} &
  \multicolumn{1}{c|}{\checkmark} &
  \multicolumn{1}{c|}{--} &
  \multicolumn{1}{c|}{--} &
  \multicolumn{1}{c|}{--} &
  -- \\ \cline{2-8} 
 &
  \begin{tabular}[c]{@{}l@{}}adaptive\\ monitor\end{tabular} &
  LEO~\cite{stillger2001leo}, ASC~\cite{haas2005automated}, PPQO~\cite{bizarro2008progressive} &
  \multicolumn{1}{c|}{--} &
  \multicolumn{1}{c|}{\checkmark} &
  \multicolumn{1}{c|}{--} &
  \multicolumn{1}{c|}{--} &
  -- \\ \hline
\multirow{6}{*}{relation-based} &
  \multirow{2}{*}{\begin{tabular}[c]{@{}l@{}}tuple\\ routing\end{tabular}} &
  \begin{tabular}[c]{@{}l@{}}Eddies~\cite{avnur2000eddies}, SteM~\cite{raman2003using}, HPE~\cite{tzoumas2010sharing},\\ CACQ~\cite{madden2002continuously}, STAIRs~\cite{deshpande2004lifting}, POLAR~\cite{justen2024polar}\end{tabular} &
  \multicolumn{1}{c|}{--} &
  \multicolumn{1}{c|}{--} &
  \multicolumn{1}{c|}{\checkmark} &
  \multicolumn{1}{c|}{--} &
  -- \\ \cline{3-8} 
 &
   &
  SHARP~\cite{bizarro2006adaptive}, ARJ~\cite{li2006adaptively} &
  \multicolumn{1}{c|}{--} &
  \multicolumn{1}{c|}{\checkmark} &
  \multicolumn{1}{c|}{\checkmark} &
  \multicolumn{1}{c|}{--} &
  -- \\ \cline{2-8} 
 &
  \multirow{3}{*}{\begin{tabular}[c]{@{}l@{}}plan\\ switching\end{tabular}} &
  Rio~\cite{babu2005proactive}, CAPE~\cite{zhu2004dynamic} &
  \multicolumn{1}{c|}{--} &
  \multicolumn{1}{c|}{\checkmark} &
  \multicolumn{1}{c|}{--} &
  \multicolumn{1}{c|}{--} &
  -- \\ \cline{3-8} 
 &
   &
  ADP~\cite{ives2004adapting} &
  \multicolumn{1}{c|}{--} &
  \multicolumn{1}{c|}{\checkmark} &
  \multicolumn{1}{c|}{\checkmark} &
  \multicolumn{1}{c|}{--} &
  -- \\ \cline{3-8} 
 &
   &
  Cuttlefish~\cite{kaftan2018cuttlefish}, SkinnerDB~\cite{trummer2021skinnerdb} &
  \multicolumn{1}{c|}{--} &
  \multicolumn{1}{c|}{--} &
  \multicolumn{1}{c|}{--} &
  \multicolumn{1}{c|}{\checkmark} &
  -- \\ \cline{2-8} 
 &
  \begin{tabular}[c]{@{}l@{}}tuple\\ sampling\end{tabular} &
  A-Greedy~\cite{babu2004adaptive} &
  \multicolumn{1}{c|}{--} &
  \multicolumn{1}{c|}{\checkmark} &
  \multicolumn{1}{c|}{--} &
  \multicolumn{1}{c|}{--} &
  -- \\ \hline
\multirow{4}{*}{plan-based} &
  \begin{tabular}[c]{@{}l@{}}checkpoints\\ checking\end{tabular} &
  \begin{tabular}[c]{@{}l@{}}Dynamic Reoptimization~\cite{kabra1998efficient},\\ POP~\cite{markl2004robust}\end{tabular} &
  \multicolumn{1}{c|}{--} &
  \multicolumn{1}{c|}{\checkmark} &
  \multicolumn{1}{c|}{--} &
  \multicolumn{1}{c|}{--} &
  \checkmark \\ \cline{2-8} 
 &
  \begin{tabular}[c]{@{}l@{}}fragments\\ checking\end{tabular} &
  SIT~\cite{bruno2002exploiting}, IEF~\cite{neumann2013taking} &
  \multicolumn{1}{c|}{--} &
  \multicolumn{1}{c|}{\checkmark} &
  \multicolumn{1}{c|}{--} &
  \multicolumn{1}{c|}{--} &
  \checkmark \\ \cline{2-8} 
 &
  \multirow{2}{*}{\begin{tabular}[c]{@{}l@{}}sub-plans\\ splitting\end{tabular}} &
  DYNO~\cite{karanasos2014dynamically}, Carac~\cite{herlihy2024adaptive} &
  \multicolumn{1}{c|}{--} &
  \multicolumn{1}{c|}{\checkmark} &
  \multicolumn{1}{c|}{--} &
  \multicolumn{1}{c|}{--} &
  \checkmark \\ \cline{3-8} 
 &
   &
  QuerySplit~\cite{zhao2023efficient} &
  \multicolumn{1}{c|}{--} &
  \multicolumn{1}{c|}{\checkmark} &
  \multicolumn{1}{c|}{--} &
  \multicolumn{1}{c|}{\checkmark\footnotemark} &
  \checkmark \\ \hline
\end{tabular}
\end{footnotesize}
\vspace{-0.5cm}
\end{table*}

The previous section gave an overview of different AQP strategies. In particular, we explained the idea behind plan-based AQP and presented the advantages and limitations of the SOTA plan-based AQP, QuerySplit.

In this section, we provide a modular design for plan-based AQP that subsumes the design supplied in QuerySplit.
To address our research questions concerning the performance impact of each AQP-related component, as outlined in Section~\ref{section:introduction}, we first describe the unified module-level representation for the existing AQP methods in Section~\ref{design:components}. This understanding enables us to identify which modules need to be modified and to help assess the tradeoff between implementation effort and performance gains. Then, based on the module-level representation of the SOTA work, QuerySplit, we extend it to support the module-level isolation measurements in Section~\ref{design:postgres}. Lastly, to examine the applicability of the SOTA plan-based AQP to in-memory DBMSs and overcome their inherent design flaws, we propose our module-level design on top of DuckDB in Section~\ref{design:duckdb}.

\subsection{Unified Module-level Representation of AQP}\label{design:components}

In this section, we describe the unified module-level representation, which highlights the components that require modification. 
Table~\ref{tab:unified_summary} provides a summary of the modules affected by each AQP approach. It is interesting to observe that different AQP strategies can modify or introduce the same modules. For example, a monitor component contributes to an inter-query AQP strategy (e.g., LEO, ASC, and PPQO), an intra-query relation-based AQP strategy (e.g., Rio, CAPE, and A-Greedy), or an intra-query plan-based AQP strategy (e.g., SIT, IEF, and DYNO). Additionally, a single AQP strategy can result in effort being spread across multiple modules. For example, almost all plan-based AQP methods require a plan splitter and a monitor.

\begin{figure}[t]
    \centering
    \includegraphics[width=.8\columnwidth]{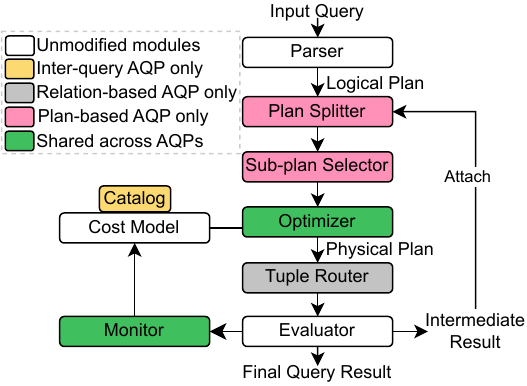}
    \caption{A unified module-level representation for AQP. The white box represents the unmodified modules from the traditional DBMS, while the colored boxes denote the AQP-related modules that are newly introduced or require modification.}
    \label{fig:3_unified_arch}
\end{figure}

The inconsistency between the AQP strategy and the module-level architectural effort prompts us to ask: What is the role of each module in the AQP approach? What is their impact on performance?

To address these questions, a unified representation is proposed (cf. Figure~\ref{fig:3_unified_arch}) that enables performance analysis in a modular manner by allowing the addition or modification of components based on the traditional database architecture skeleton. In the unified representation diagram, white rectangles represent the modules inherent from the traditional DBMS architecture without any changes, including the parser, cost model, and evaluator. The monitor module, which is used for collecting and updating refined statistics, is commonly employed across various AQP strategies.
The adaptive catalog does not alter the functionality of the cost model; instead, it adapts the data distribution based on past query execution. The existing literature has only utilized it in inter-query AQP, but it should also apply to intra-query AQP. The plan splitter and sub-plan selector only work on the plan-based AQP, while the tuple router only works with the relation-based AQP. The learning-based AQP methods, e.g., Cuttlefish~\cite{kaftan2018cuttlefish} and SkinnerDB~\cite{trummer2021skinnerdb}, remove the traditional optimizer and cost model, and use a learned-based optimizer instead. 
In addition, as described in Section~\ref{preliminaries:query_split}, QuerySplit adds join reordering rules by the sub-plan selector, which acts like an optimizer. 
As suggested by the hints from QuerySplit, classical heuristic optimization rules can be combined with AQP, and, in theory, are not constrained by the types of the AQP category.



To address the research questions of the performance impact of each module that we asked in Section~\ref{section:introduction}, we evaluate the plan-based AQP against the traditional architecture on both in-memory and on-disk DBMS in a systematic manner. There is a set of plan-based AQP methods as shown in Table~\ref{tab:unified_summary}, where we focus on the SOTA work, QuerySplit, because the module it modifies - \textcircled{\small{1}} the \textbf{plan splitter}, \textcircled{\small{2}} the \textbf{sub-plan selector}, and \textcircled{\small{3}} the \textbf{monitor} - covers the key components affected by other approaches.
This observation guides our performance analysis, which is based on the principle of modular isolation.

\begin{figure}[t]
    \centering
    \includegraphics[width=.8\columnwidth]{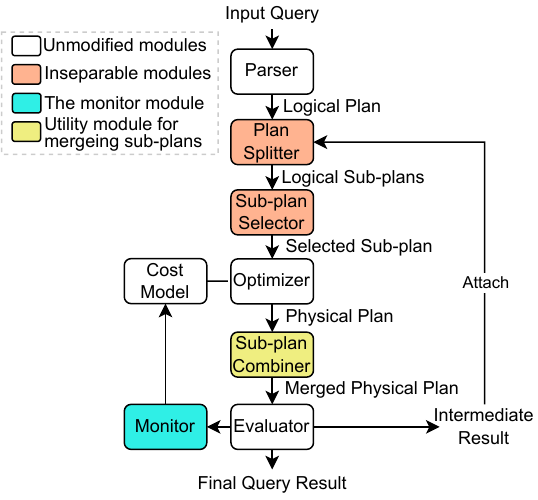}
    \caption{The modular design for plan-based AQP.}
    \label{fig:3_impl}
\end{figure}

\footnotetext[2]{QuerySplit has a sub-plan selector, which has the effect of join reordering.}

\subsection{Deployment in PostgreSQL}\label{design:postgres}

Based on the description of the module-level representation of QuerySplit in Section~\ref{design:components}, this section describes our extension, \textbf{AQP-PostgreSQL}, which supports the options of enabling or disabling the plan splitter, sub-plan selector, and monitor (cf. Figure~\ref{fig:3_impl}).
Because of the design of QuerySplit, the plan splitter and sub-plan selector modules are inherently inseparable. We achieve the same logical effect as isolating these modules by introducing an additional utility module.

To facilitate the description of the impact of each module, we first introduce the implementation of disabling the monitor module, then describe the plan splitter module, and finally analyze the sub-plan selector module.

\subsubsection{Enable/disable the monitor}\label{impl:pg_update_card}
AQP-PostgreSQL follows the \texttt{VACUUM} and \texttt{ANALYZE} processes from PostgreSQL to collect and update the cardinality of the intermediate result by providing the corresponding relation \texttt{OID}. 
It is straightforward to disable the monitor module in AQP-PostgreSQL (cf. Figure~\ref{fig:3_impl}); the temporary table containing the intermediate result must be attached to the remaining query plan immediately, without any further analysis or cardinality update.

\subsubsection{Enable/disable the plan splitter}\label{impl:pg_merge_sub_plan}

\begin{figure}[!t]
    \centering
    \includegraphics[width=.62\columnwidth]{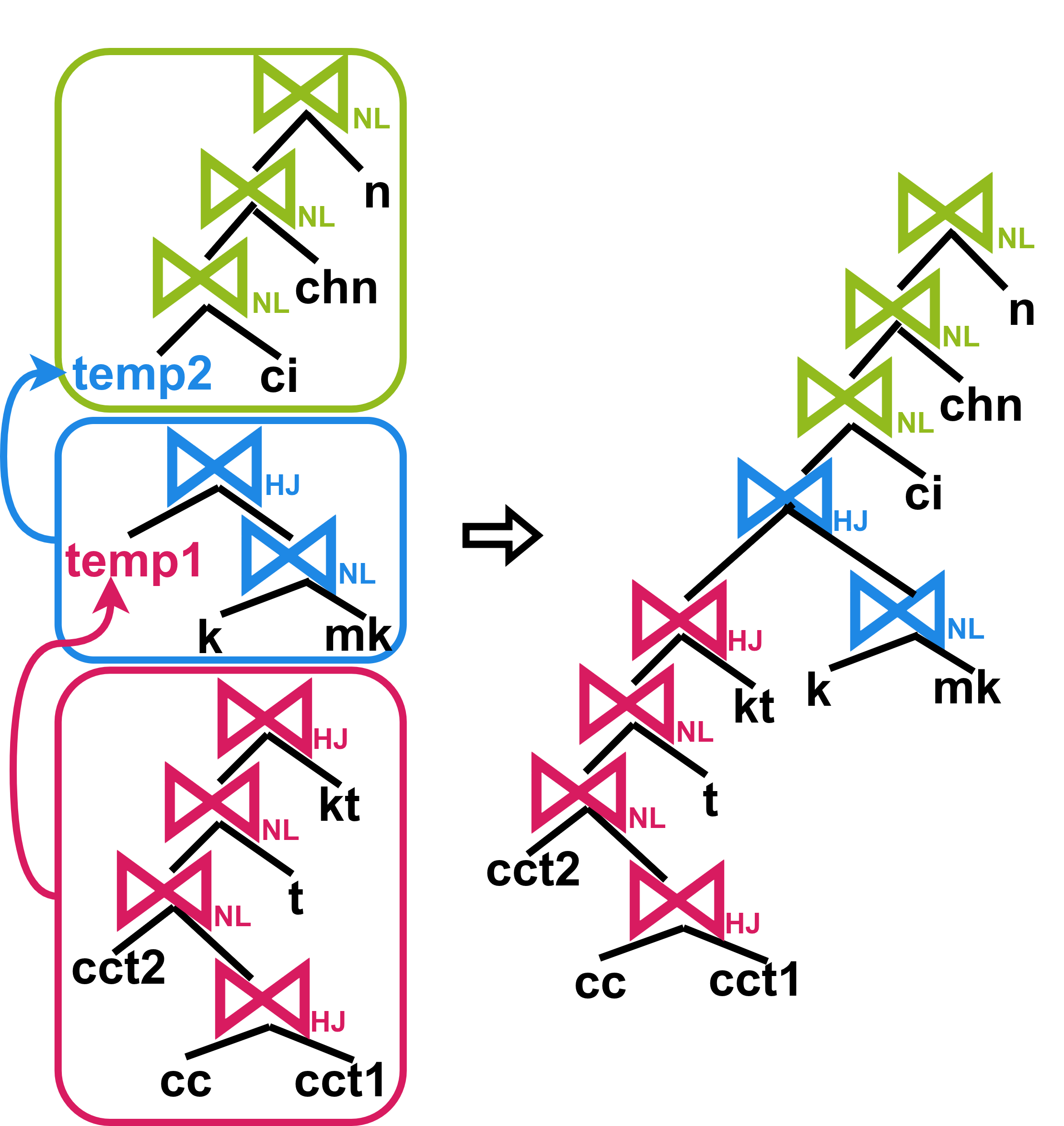}
    \caption{In the example of query 20c.sql from JOB, we merge the sub-plans in order from bottom to top.}
    \label{fig:7_rq3_example}
\end{figure}

The difference between enabling and disabling the plan splitter is whether to execute sub-plans one by one or execute the complete plan directly with the same plan order. AQP-PostgreSQL provides the option to merge all sub-plans into a complete plan. 
We consider our running example from Section~\ref{preliminaries:query_split}, query \texttt{20c.sql} from JOB, to explain how it works.
QuerySplit splits the plan of this query into three sub-plans (cf. the right part in Figure~\ref{fig:2_example}), and the first two sub-plans are stored in a temporary table. We introduce a utility module (cf. Figure~\ref{fig:3_impl}) to merge the three sub-plans into a complete plan by replacing the ``temporary'' table with the corresponding sub-plan (cf. Figure~\ref{fig:7_rq3_example}).
In detail, for each round, AQP-PostgreSQL creates a temporary table to store the intermediate result of each sub-plan and keeps a copy of the sub-plan for merging.
When the second sub-plan comes, AQP-PostgreSQL adapts the corresponding table and column indexes and merges the two sub-plans.
The merged plan is then sent to the executor, and the new intermediate result is obtained. AQP-PostgreSQL repeats this process until it collects the last sub-plan.

\subsubsection{Enable/disable the sub-plan selector}\label{impl:pg_select_sub_plan}

As we described in Section~\ref{preliminaries:query_split}, QuerySplit splits the plan and selects the sub-plans on top of the DAG representation, making it inherently impossible to disable the sub-plan selector module (cf. Figure~\ref{fig:3_impl}). However, we measure the effect of the sub-plan selector module by comparing the vanilla plan to the merged plan without updating cardinality, where the only enabled module is the sub-plan selector.

\subsection{Deployment in DuckDB}\label{design:duckdb}


Based on the module-level study in Section~\ref{design:components}, we propose a new design, AQP-DuckDB, on top of a in-memory DBMS, DuckDB. Also, the extension of enabling/disabling the plan splitter, sub-plan selector, and monitor (cf. Figure~\ref{fig:3_impl}) is supported similarly to AQP-PostgreSQL in Section~\ref{design:postgres}.

At a high level, AQP-DuckDB follows the pipeline from QuerySplit but replaces the DAG generator module with a join reorder module to avoid its drawbacks. 
As described in Section~\ref{preliminaries:query_split}, the DAG representation has limitations, including foreign key constraints and no support for non-SPJ queries. Additionally, it incurs the performance overhead of iterating sub-graphs (sub-plans). 
AQP-DuckDB inherits the insight of prioritizing the smallest possible sub-plan by reordering the relations from the smallest to the largest in the plan tree (which actually changes the join order). However, it supports both SPJ queries and non-SPJ queries natively by splitting based on the logical plan tree. At the same time, AQP-DuckDB has a new split strategy, which splits at each \texttt{FILTER} and \texttt{JOIN} operators, thereby eliminating the restrictions imposed by foreign key constraints.

\subsubsection{Implementation of Join Reorder}\label{impl:duckdb_join_reorder}
AQP-DuckDB has a join reorder module before the plan splitting phase that applies a heuristic by executing the smallest relation first and canonicalizing the \texttt{CROSS\_PRODUCT} nodes. In this module, AQP-DuckDB sorts relations from largest to smallest based on their row count. Note that this module treats the \texttt{FILTER} node at the same level as the relation node and participates in the sorting together if it only has one relation. This is because usually such \texttt{FILTER} node is bound to its relation child in the plan tree. Then we fill in other types of operator nodes and reschedule the position of the \texttt{CROSS\_PRODUCT} nodes and \texttt{JOIN} nodes to make sure the join order otimizer can simplify every \texttt{CROSS\_PRODUCT} for each sub-plan. AQP-DuckDB has the option to disable this module and canonicalize the \texttt{CROSS\_PRODUCT} nodes without plan reordering.

\subsubsection{Implementation of Plan Splitter and Sub-plan Selector}\label{impl:duckdb_plan_split_select}
After the join reorder module, AQP-DuckDB splits the plan into several sub-plans at each filter and join operators, collecting the necessary expression information for each sub-plan. Then, the sub-plan selector module selects the sub-plan to be executed in the current round in a bottom-up order. Alternatively, executing the join order optimizer exposes pipeline breaker points (i.e., the build phase of a hash join), which can improve the plan-splitting strategy. Once the cardinalities are updated, the rest of the plan is reoptimized and split again.

\subsubsection{Implementation of Sub-plan Result Feeding Back}\label{impl:duckdb_feedback}
Once AQP-DuckDB executes the sub-plan, the intermediate result will be stored in memory and then will be attached to the remaining logical plan. Here, it also collects the cardinality of the intermediate result and updates the cost model with this information. The intermediate result will be allocated to a new table index, which means AQP-DuckDB needs to update the corresponding index and expressions.


\subsubsection{Enable/disable the monitor}\label{impl:duckdb_update_card}
To better understand the performance effect of the monitor module, AQP-DuckDB can disable it by using the cardinality estimated by vanilla DuckDB without further analyzing the intermediate result, or to enable it by refining the cardinality based on the execution of the previous sub-plan (cf. Figure~\ref{fig:3_impl}). The process of updating the refined cardinality is described in Section~\ref{impl:duckdb_feedback}.
For DuckDB's estimated cardinality of each node, we choose the largest one of its child nodes.

\subsubsection{Enable/disable the plan splitter}\label{impl:duckdb_merge_sub_plan}
Similarly to Section~\ref{impl:pg_merge_sub_plan}, AQP-DuckDB provides the option to merge all sub-plans into a whole plan. The primary difference between AQP-DuckDB and AQP-PostgreSQL is that the former retains intermediate results in memory, whereas the latter creates a temporary table to store them. 
As a result, AQP-DuckDB keeps a copy of each sub-plan in memory after the join order optimization. Then, it reverts the table and column indexes and merges them to the remaining plan in-place (cf. Figure~\ref{fig:3_impl}). We use depth-first search to detect the attachment point on top of the bushy plan generated from DuckDB's optimizer. Finally, AQP-DuckDB builds a complete plan with the same execution order as the sub-plans.

\subsubsection{Enable/disable the sub-plan selector}\label{impl:duckdb_select_sub_plan}
AQP-DuckDB decouples the sub-plan selector and the join reorder as described in Section~\ref{impl:duckdb_plan_split_select}. 
Its selection strategy is to pick the sub-plans in a bottom-up fashion in the plan tree. Selecting the sub-plan with the least estimated cardinality in each round is based on the join reorder module described in Section~\ref{impl:duckdb_join_reorder}. To analyze the performance effect of the sub-plan selector module, we employ the same strategy as AQP-PostgreSQL in Section~\ref{impl:pg_select_sub_plan}, by disabling the monitor and plan splitter, and comparing it with vanilla DuckDB.

\section{Experimental Setup}\label{section:setup}

In this section, we describe the general setup of the evaluation, including the prototype, testbed, workloads, and the metrics we measure and evaluate.

\smartpara{Prototype}\label{setup:prototype}
We exploited and reproduced the open-source QuerySplit project \cite{query_split_code} in PostgreSQL, adapting to the Linux Operating System and extending it with the performance breakdown and module-level analysis (with 2100 lines of C code). We also apply the idea of plan-based AQP into DuckDB (v0.10.1 and v1.3.2, separately) with a new design (with 8880 lines of C++ code), requiring minimal changes to the existing DuckDB codebase. 
To compare with the relation-based AQP, we reproduced the open-source Polar project~\cite{justen2024polar}, which is based on DuckDB. For a fair comparison, we ported our implementation to the same DuckDB version (v0.6.1).
We open-source our implementation and experimental materials.

\smartpara{Hardware} We conducted all experiments on a \texttt{x86/64} Dell server with a single Intel(R) Xeon(R) E-2236 CPU at 3.40GHz, 
64 GB DDR4 RAM at 2666 MHz, 1 $\times$ 931.5 GB SATA HDD (6GB/s).

\smartpara{Software} We compiled the source code with gcc-11.4.0 on Ubuntu 22.04.5 LTS. We use the same PostgreSQL version as QuerySplit (12.3)~\cite{postgres_code}, 
For DuckDB, we choose the latest release version at the time we start this project (0.10.1)~\cite{duckdb_code}, and a more recent version (1.3.2)~\cite{duckdb_132_code}. We also implemented it in the same version as the Polar~\cite{justen2024polar} project (0.6.1).

\begin{table}[]
\caption{Configurations for PostgreSQL.}\label{tab:pg_settings}
\vspace{-0.35cm}
\begin{center}
\begin{small}
\begin{tabular}{|l|l|}
\hline
\textbf{Parameters}                 & \textbf{Configurations} \\ \hline
parallel\_leader\_participation     & off                     \\ \hline
max\_parallel\_workers              & 0                       \\ \hline
max\_parallel\_workers\_per\_gather & 0                       \\ \hline
shared\_buffers                     & 512 MB                  \\ \hline
temp\_buffers                       & 2047 MB                 \\ \hline
work\_mem                           & 2047 MB                 \\ \hline
effective\_cache\_size              & 4 GB                    \\ \hline
statement\_timeout                  & 1000 s                  \\ \hline
default\_statistics\_target         & 100                     \\ \hline
\end{tabular}
\end{small}
\end{center}
\end{table}

\smartpara{Measurement tools and method} For the end-to-end time, we use the hyperfine tool \cite{hyperfine_tool} with 5 times warmup, and we consider the average of 10 runs. 
For the execution time, we use the timer function with the high-resolution clock in C++ \cite{high_resolution_clock}. 
When measuring the vanilla PostgreSQL and the AQP-PostgreSQL, we use the same configurations as the ones employed in the QuerySplit paper~\cite{zhao2023efficient} (cf. Table \ref{tab:pg_settings}) and collect the statistics by running `\texttt{ANALYZE;}` before measuring the performance. When measuring the vanilla DuckDB, Polar, and the AQP-DuckDB, we run them on a single thread\footnote{We use the same configurations as the SOTA (QuerySplit), which uses one thread, to have a fair comparison.} by setting the `\texttt{maximum\_threads}` to 1. The remaining settings are the default configurations.

\subsection{Workloads}\label{setup:workloads}
We use two workloads for our evaluation: (1) the JOB with the IMDB dataset~\cite{job_benchmark} and (2) the DSB~\cite{dsb_benchmark}.
For our scalability evaluation, we consider two different scale factors for the DSB benchmark: 10 and 100.

In vanilla DuckDB, Polar, and AQP-DuckDB, we support the whole JOB benchmark with a total of 113 queries and the entire DSB benchmark with a total of 58 queries of the SPJ and aggregation categories.\footnote{Except for  `\texttt{query072}` where there is an open issue in DuckDB~\cite{duckdb_072_issue}}
In Postgres, only 63 queries from the JOB benchmark and 14 queries of the SPJ category from the DSB benchmark are supported.\footnote{This is due to the engineering issues in the available implementation of QuerySplit (e.g., segmentation fault and attributes with wrong types).}

\subsection{Evaluation Metrics}\label{setup:metrics}
We verify the correctness and evaluate the running time as our main metric. For correctness verification, we generate the golden data from the vanilla version and compare the results of each query with those of the AQP version. For running time, we evaluate the sum of each benchmark. 
Because the number of queries supported by AQP-DuckDB and AQP-PostgreSQL differs, we include a column in the middle of each figure to highlight the performance of queries that are supported in both AQP-DuckDB and AQP-PostgreSQL, labeled as \textbf{DuckDB (common)}. 
We then present the query-by-query performance distribution to clarify whether the performance improvement or degradation stems from individual queries or if there is a general trend among them.

\section{Experimental Evaluation}\label{section:eval}

\begin{figure*}[t]
  \centering
  \begin{subfigure}[t]{0.19\textwidth}
    \includegraphics[height=4cm]{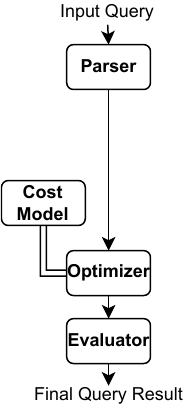}
    \caption{Vanilla DBMS  \\ (RQ1a, and RQ3a).}\label{fig:5_1_vanilla}
  \end{subfigure}
  \hfill
  \begin{subfigure}[t]{.21\textwidth}
    \includegraphics[height=4cm]{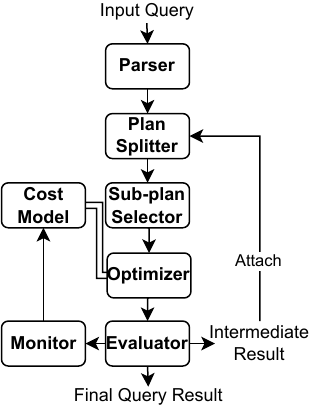}
    \caption{AQP \\ (RQ1b, RQ2b, and RQ4b).}\label{fig:5_2_AQP}
  \end{subfigure}
  \hfill
  \begin{subfigure}[t]{0.19\textwidth}
    \includegraphics[height=4cm]{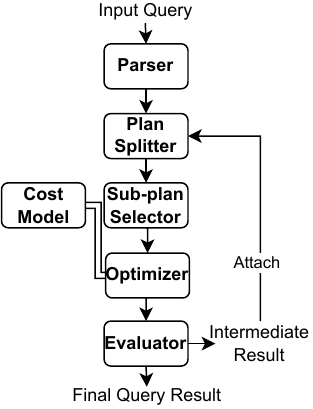}
    \caption{AQP w/o monitor\\ (RQ2a).}\label{fig:5_3_AQP_wo_stat}
  \end{subfigure}
  \hfill
  \begin{subfigure}[t]{0.19\textwidth}
    \includegraphics[height=4cm]{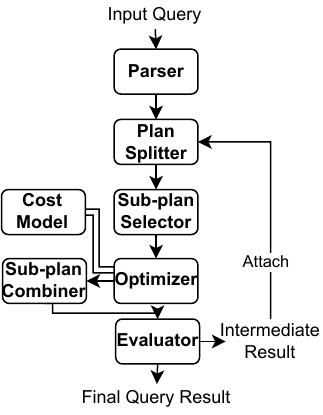}
    \caption{Vanilla w/ AQP's plan \\ (RQ3b).}\label{fig:5_4_order}
  \end{subfigure}
  \hfill
  \begin{subfigure}[t]{0.19\textwidth}
    \includegraphics[height=4cm]{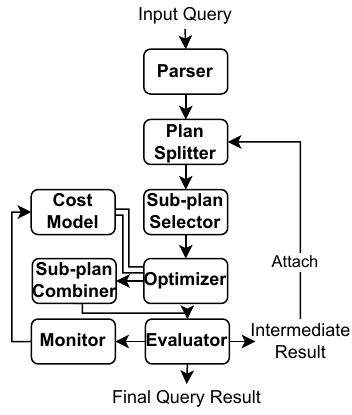}
    \caption{AQP w/o splitter \\ (RQ4a).}\label{fig:5_5_wo_split}
  \end{subfigure}
  \caption{The comparison of the traditional architecture with different plan-based AQP alternatives. The related research question and its variant are reported in parentheses. For example, RQ3b refers to the second variant used in research question 3. }
    \label{fig:2_arch_compare}
\end{figure*}

Our study conducts an experimental evaluation for the three key modules of the plan-based AQP strategy, as we mentioned in Section~\ref{design:components}: \textcircled{\small{1}} \textbf{plan splitter}, \textcircled{\small{2}} \textbf{sub-plan selector}, and \textcircled{\small{3}} \textbf{monitor}.
There are dependencies between these modules. In each round, plan-based AQP first splits the plan into sub-plans, then selects the sub-plan with the least estimated cardinality to execute. After optimizing and executing the chosen sub-plan, the monitor collects the ``actual'' cardinality and refines the cost model for the subsequent rounds.

To separate the performance effects of the above three modules, we measured the performance by enabling/disabling each one. 
At the end, we measured the performance comparison of Polar and AQP-DuckDB (0.6.1).
The following research questions drive the evaluation: 
\begin{itemize}
    \item RQ1: Is plan-based AQP beneficial for PostgreSQL and DuckDB, as a representative for on-disk and in-memory DBMSs, respectively (Section~\ref{eval:overall})?
    \item RQ2: To what extent does the monitor component improve the performance (Section~\ref{eval:updating_stats})?
    \item RQ3: What is the impact of sub-plan selector on the performance (Section~\ref{eval:join_order})?
    \item RQ4: How much is the overhead associated with the plan-splitter component (Section~\ref{eval:splitter})?
    \item RQ5: How good is the plan-based AQP compared to the relation-based AQP (Section~\ref{eval:relation-based})?
\end{itemize}


\begin{figure*}[t]
    \centering
    \begin{subfigure}[t]{\textwidth}
    \vspace{-0.3cm}
    \includegraphics[width=\linewidth]{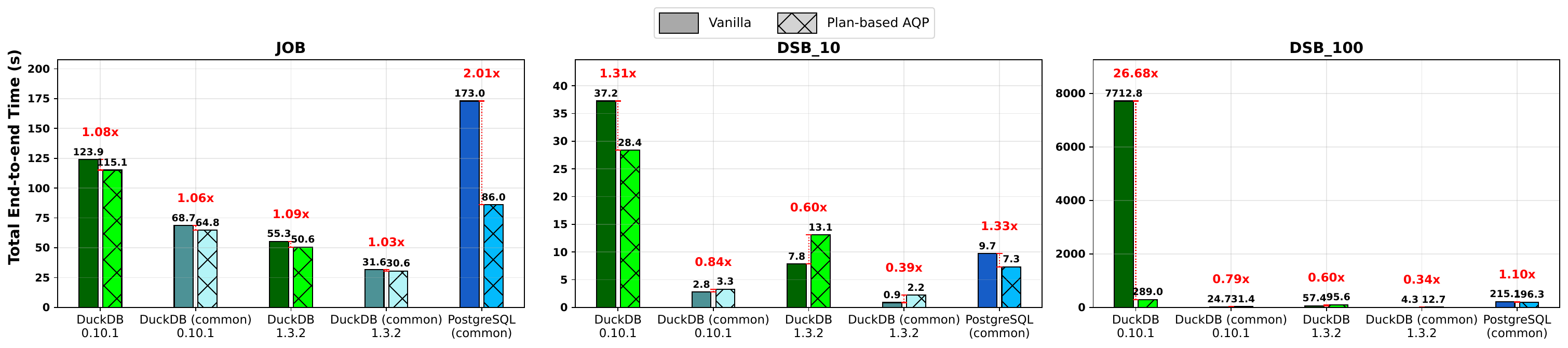}
    \caption{Total end-to-end time comparison for different benchmarks.}
    \label{fig:q1_fig_common}
    \end{subfigure}
    \centering
    \begin{subfigure}[t]{\textwidth}
    \includegraphics[width=\linewidth]{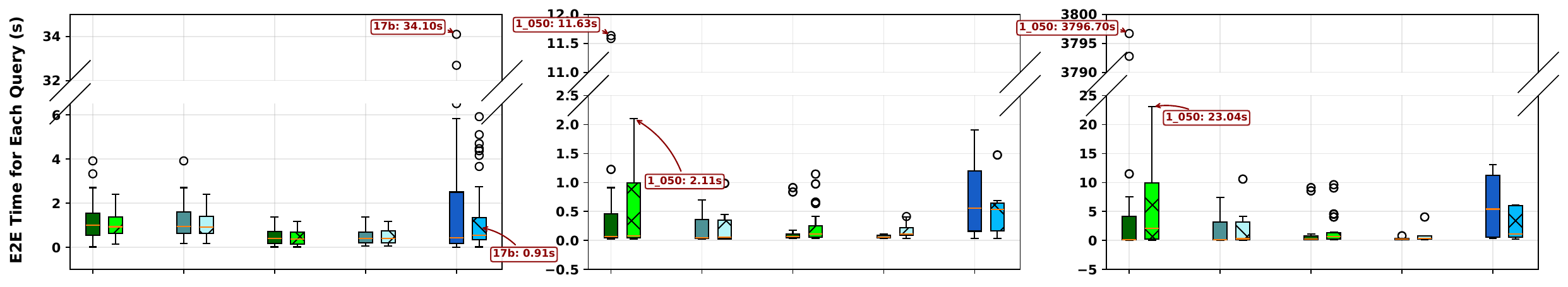}
    \caption{End-to-end time distributions for different benchmarks.}
    \label{fig:box_qbq_q1_fig}
    \end{subfigure}
    \caption{Performance comparison between the vanilla DBMS and the plan-based AQP strategy.}
    \label{fig:q1_fig}
    \vspace{-0.5cm}
\end{figure*}



\subsection{Overall Evaluation}\label{eval:overall}
In this section, we aim to answer \textbf{RQ1} and have an intuitive understanding of the overall performance impact of the plan-based AQP strategy. We measure vanilla DuckDB and PostgreSQL as the baseline (cf. Figure~\ref{fig:5_1_vanilla}), and AQP-DuckDB and AQP-PostgreSQL as the variants (cf. Figure~\ref{fig:5_2_AQP}) for both the JOB and DSB benchmarks.

\smartpara{Overall} Figure~\ref{fig:q1_fig_common} shows the total end-to-end time comparison results, based on which we make the following observations. 
First, the plan-based AQP strategy, implemented on top of both DuckDB and PostgreSQL, achieves end-to-end speedup for the JOB, but gets a slowdown for most cases of the DSB workload. Second, AQP-DuckDB experiences a slowdown in the DSB benchmark for common queries. 
Third, we observe that AQP-DuckDB (0.10.1), with the DSB benchmark using a scale factor of 100, yields a significant performance improvement, which is up to $25.68\times$ faster than vanilla DuckDB.

Based on the former evaluation results, it is not clear what the contributions of performance improvement are for each of the queries. 
Is it due to some extreme queries that have achieved significant speedup, or have almost all queries improved their performance?

\smartpara{Drilldown}
To address this, we analyze the performance distribution of the two benchmarks by using box plots (cf. Figure~\ref{fig:box_qbq_q1_fig}). 
In each sub-figure, the box plot shows the end-to-end time distribution of the corresponding system, with the same color and pattern aligning with the total time comparison (cf. Figure~\ref{fig:q1_fig_common}), where each point represents the end-to-end time of a single query.
By analyzing the end-to-end time distribution, we observe that long-running queries play a crucial role in achieving performance improvements, particularly for PostgreSQL in the JOB benchmark and DuckDB in the DSB benchmark. However, for DSB, both versions of DuckDB experience a slowdown with plan-based AQP for most queries.

\smartpara{Case Study}
By checking the end-to-end time of each query, we notice several outliers, e.g., query \texttt{17b.sql} from the JOB benchmark in PostgreSQL and query \texttt{1\_050.sql} from the DSB benchmark in DuckDB (0.10.1) with scale factors of 10 and 100, respectively, have been significantly improved. For each box plot, we only label the most significant outlier to prevent occlusion.
In fact, even if we remove the outlier from the PostgreSQL on the JOB benchmark, we can still find an improvement. However, this is not the case for DuckDB (0.10.1) on the DSB benchmark.
For example, in the DSB benchmark with a scale factor of 100, queries '1\_050' and '2\_050' each spend around $3,795$ seconds in vanilla DuckDB, but only $23$ seconds in AQP-DuckDB. After removing these two groups of outliers, the performance of AQP-DuckDB drops to $0.53 \times$ that of vanilla DuckDB, which means a considerable slowdown.

\smartpara{Conclusion} Although the overall performance of the two systems on the JOB benchmark has improved, the reasons behind this are inconsistent. Notably, the plan-based AQP on top of DuckDB for the DSB benchmark experiences a significant slowdown. The actual source of speedup or slowdown for both PostgreSQL and DuckDB is yet to be determined. 

\begin{figure*}[!t]
    \centering
    \begin{subfigure}[t]{\textwidth}
    \includegraphics[width=\textwidth]{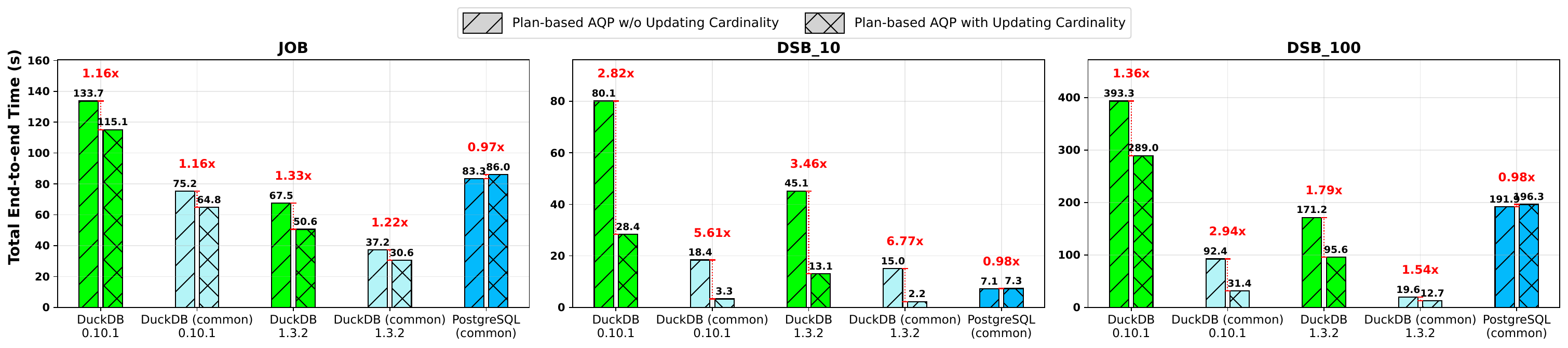}
    \caption{Total end-to-end time for different benchmarks.}
    \label{fig:q2_fig_common}
    \end{subfigure}
    \centering
    \begin{subfigure}[t]{\textwidth}
    \centering
    \includegraphics[width=\textwidth]{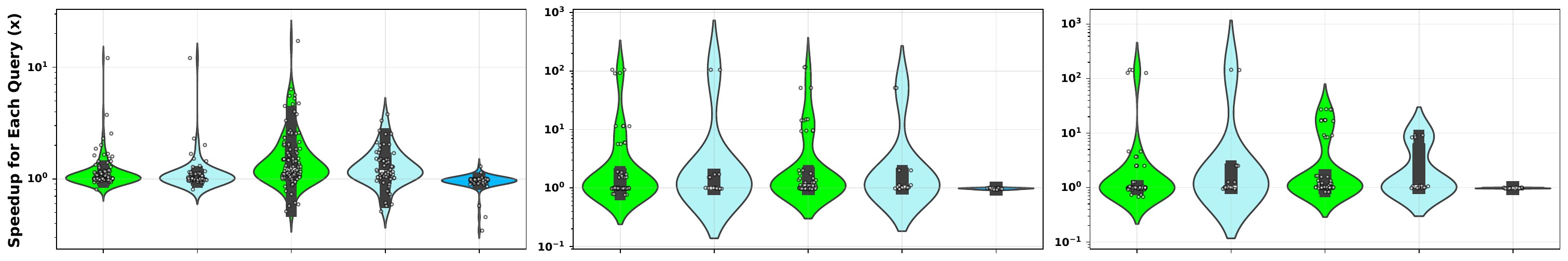}
    \caption{Query-by-query performance improvement distributions for different benchmarks.}
    \label{fig:qbq_compare_q2_fig}
    \end{subfigure}
    \caption{Performance comparison with/without updating cardinality.}
    \label{fig:q2_fig}
    \vspace{-0.2cm}
\end{figure*}

\subsection{Performance Effect of the Monitor}\label{eval:updating_stats}


In this section, we aim to answer \textbf{RQ2} by analyzing the impact of the \textbf{monitor} module, which is mainly responsible for analyzing and updating the cardinality. To compare the performance with or without the monitor module, we use the plan-based AQP without updating cardinality (as described in Section~\ref{impl:duckdb_update_card} and Section~\ref{impl:pg_update_card}) as the baseline (cf. Figure~\ref{fig:5_3_AQP_wo_stat}), and the one with cardinality refined as the variants (cf. Figure~\ref{fig:5_2_AQP}) for both the JOB and DSB benchmarks.

\smartpara{Overall}
For both benchmarks, AQP-DuckDB with updated cardinality achieves a significant speedup by updating the refined cardinality. 
However, AQP-PostgreSQL does not show a clear improvement from the monitor module; to make matters worse, it shows a slight performance slowdown for $0.97 \times$, $0.98 \times$, and $0.98 \times$ (cf. Figure~\ref{fig:q2_fig_common}). 

\smartpara{Drilldown} Similarly to what we discussed in Section~\ref{eval:overall}, the total end-to-end time figure cannot tell us the performance improvement contribution from each query. We first analyzed the box plot to compare the AQP with/without the monitor module, but we did not find any obvious outliers (eight speedups or slowdowns). This indicates that the performance impact of the monitor module is not caused by individual extreme queries or long-running queries, as observed in Section~\ref{eval:overall}, but rather is common to almost all queries. 
However, it is unclear whether there is a large number of queries whose performance is degraded, or the overall performance is improved mainly because of the other queries.

To address this, we further examine the distribution of query-by-query improvement (cf. Figure~\ref{fig:qbq_compare_q2_fig}) to determine the percentage of queries with improved and degraded performance for each system and corresponding benchmarks. For the improvement distribution (cf. Figure~\ref{fig:qbq_compare_q2_fig}), we use a violin plot on a log scale, where larger values indicate greater relative improvements in performance. However, a query with a significant performance improvement may not necessarily have a considerable running time. This is the main distinction between the performance improvement distribution (violin plot) and the query's end-to-end time distribution (box plot).

\smartpara{Distribution Study}
By checking the query-by-query performance improvement distribution (cf. Figure~\ref{fig:qbq_compare_q2_fig}), we observe that AQP-DuckDB gains performance improvement from almost all queries for both the JOB benchmark and the DSB benchmark, and only a few (less than $10 \%$) queries have a slight slowdown. However, the performance of most AQP-PostgreSQL queries in the JOB benchmark decreases, while only two queries show a slight improvement in performance. For the DSB benchmark with the two different scale factors, it does not present a performance difference when the monitor module is disabled or enabled. It is worth noting that, although queries run by AQP-PostgreSQL can, in theory, gain speedup from the refined cardinality, as mentioned in Section~\ref{preliminaries:on_disk_and_in_memory}, there is an overhead of generating a temporary table to store the intermediate result and analyzing it. This is the main reason for the performance degradation of AQP-PostgreSQL.

\smartpara{Conclusion} 
Monitoring the cardinality and using updated ones speeds up almost all queries in AQP-DuckDB. However, the monitor component slows down AQP-PostgreSQL due to the additional overhead of storing the intermediate relations.

\subsection{Performance Effect of the Sub-plan Selector}\label{eval:join_order}



\begin{figure*}[t]
    \centering
    \begin{subfigure}[t]{\textwidth}
    \includegraphics[width=\linewidth]{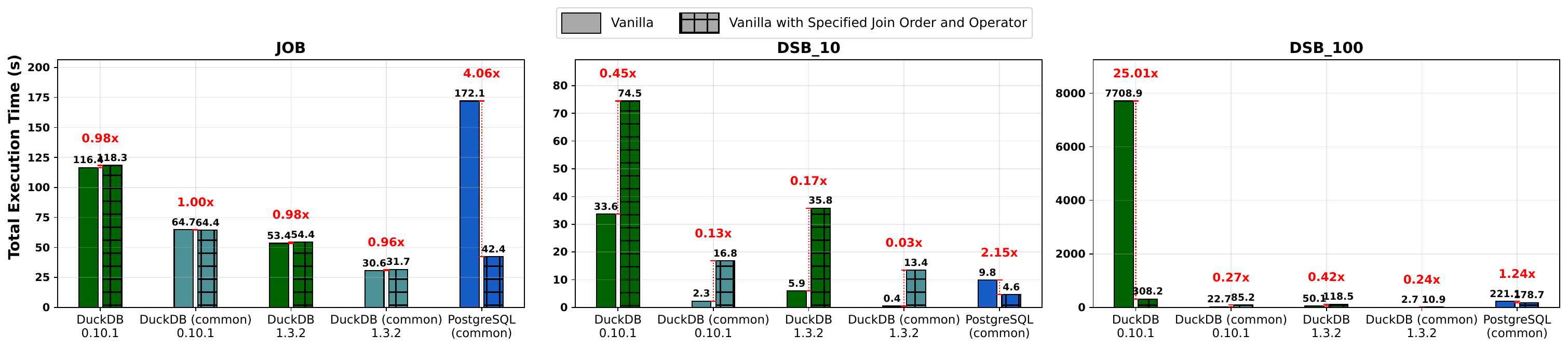}
    \caption{Total execution time for different benchmarks.}
    \label{fig:q3_fig_common}
    \end{subfigure}
    \centering
    \begin{subfigure}[t]{\textwidth}
    \includegraphics[width=\linewidth]{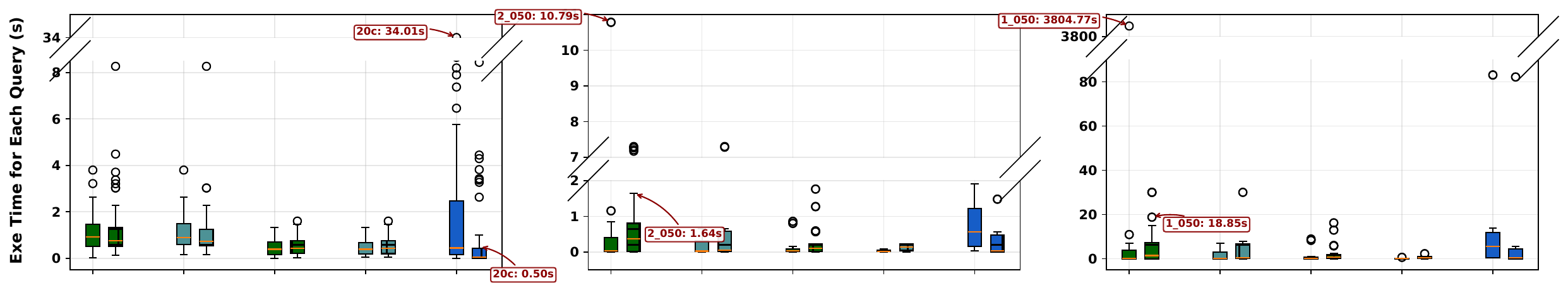}
    \caption{Execution time distributions for different benchmarks.}
    \label{fig:box_qbq_q3_fig}
    \end{subfigure}
    \caption{Performance comparison with/without specifying the join order and operators.}
    \label{fig:q3_fig}
    \vspace{-0.4cm}
\end{figure*}

In this section, we aim to answer \textbf{RQ3} by analyzing the effect of the \textbf{sub-plan selector}. 
As described in Section~\ref{preliminaries:query_split}, the sub-plan selector is an inseparable module, and we cannot disable it directly. However, as described in Section~\ref{impl:pg_select_sub_plan} and Section~\ref{impl:duckdb_select_sub_plan}, we introduce a sub-plan combiner module on top of our unified module-level representation to mitigate the effect of the plan-splitter module. We then disable the monitor module, and the only distinguished module from the vanilla DBMS is the sub-plan selector. 
For a more precise comparison and analysis, we simplify the description to vanilla DBMSs with a specified join order and operators.
Thus, we can analyze the effect of the sub-plan selector by using the vanilla DBMSs as the baseline (cf. Figure~\ref{fig:5_1_vanilla}) and comparing it with the vanilla DBMSs with a specified join order and operators as the variants (cf. Figure~\ref{fig:5_4_order}).
Since the end-to-end time of this process includes the overhead of merging sub-plans, we measure the execution time instead.

\begin{figure*}[!t]
    \centering
    \begin{subfigure}[t]{\textwidth}
    \includegraphics[width=\textwidth]{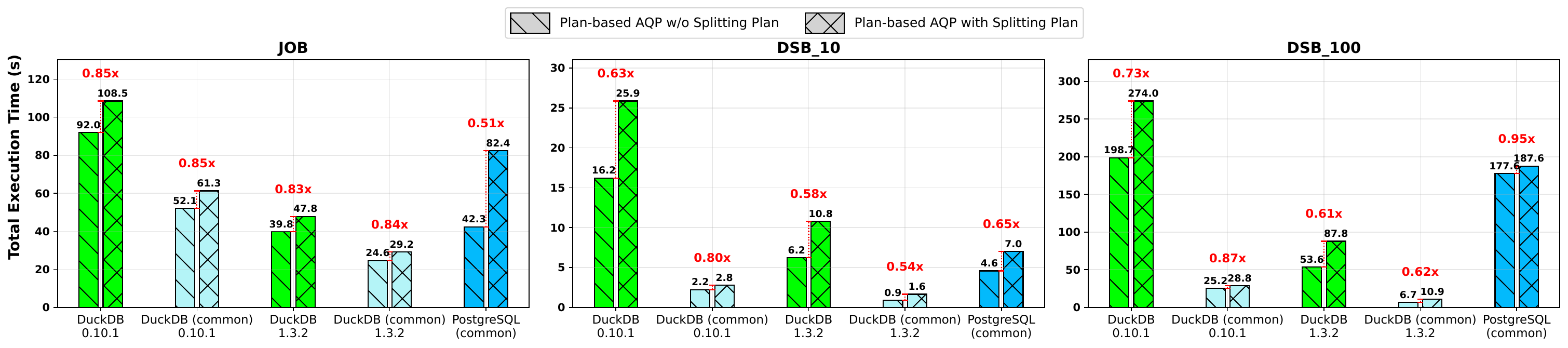}
    \caption{Total end-to-end time for different benchmarks.}
    \label{fig:q4_fig_common}
    \end{subfigure}
    \centering
    \begin{subfigure}[t]{\textwidth}
    \includegraphics[width=\textwidth]{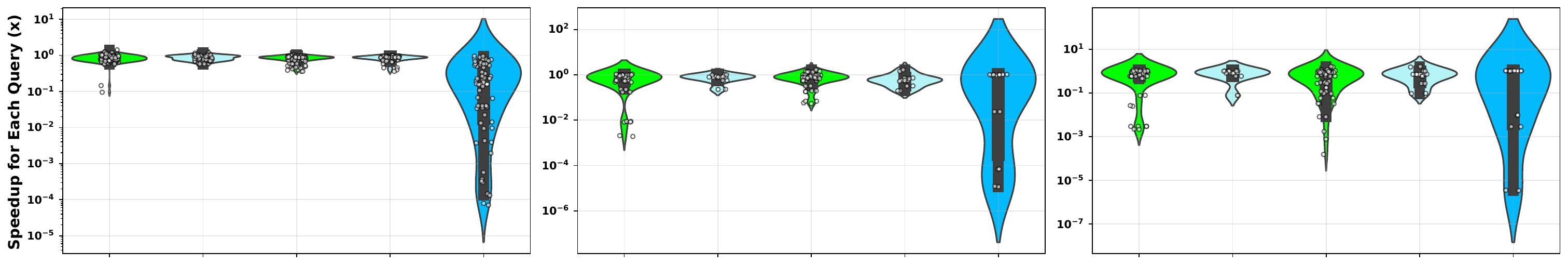}
    \caption{Query-by-query performance improvement distributions for different benchmarks.}
    \label{fig:qbq_compare_q4_fig}
    \end{subfigure}
    \caption{Performance comparison with/without splitting the plan.}
    \label{fig:q4_fig}
    \vspace{-0.4cm}
\end{figure*}

\smartpara{Overall}
The performance shown in different benchmarks on various systems is inconsistent (cf. Figure~\ref{fig:q3_fig_common}). In detail, DuckDB does not exhibit a clear speedup in the JOB benchmark; instead, it shows a slight slowdown. However, it presents significant performance degradation in the DSB benchmark with a scale factor of 10, but with dramatic improvement with a scale factor of 100 with version 0.10.1. We observe the same significant speedup from the DSB benchmark with a scale factor of 100 in DuckDB (0.10.1), as we did in Section~\ref{eval:overall}. On the other hand, PostgreSQL gains speedups for both the JOB benchmark and the DSB benchmark.

\smartpara{Drilldown}
Next, we analyze the distribution of query execution times (cf. Figure~\ref{fig:box_qbq_q3_fig}). DuckDB has long-running queries that exhibit an apparent slowdown (except for the extreme outlier queries, \texttt{1\_050.sql} and \texttt{2\_050.sql}), in both the JOB benchmark and the DSB benchmark. PostgreSQL behaves oppositely. For the long-running queries, we observe noticeable performance improvements. However, similar to the issues we encountered in Section~\ref{eval:updating_stats}, it is unclear whether there are queries that exhibit performance changes contrary to the overall performance. As a result, we further examine the distribution of query-by-query improvement (cf. Figure~\ref{fig:qbq_compare_q3_fig}).

\smartpara{Case Study}
Extreme performance improvements occur for both DuckDB and PostgreSQL in long-running queries (cf. Figure~\ref{fig:box_qbq_q3_fig}). Queries \texttt{1\_050.sql} and \texttt{2\_050.sql} show the exact same behavior as we observe in Section~\ref{eval:overall}. The significant improvement in the performance of these two long-running queries has resulted in an overall enhancement of the entire benchmark's performance. Similar to what we did in Section~\ref{eval:overall}, after removing the outliers ($3804.77$ seconds and $3789.78$ seconds for queries \texttt{1\_050.sql} and \texttt{2\_050.sql} in vanilla DuckDB (0.10.1) and $18.85 s$ and $18.84 s$ with specified join order and operators), the performance drops to $0.42\times$ that of vanilla DuckDB (0.10.1). This explains the huge performance improvement of AQP-DuckDB  (0.10.1) over the DSB benchmark in RQ1 (cf. Figure~\ref{fig:q1_fig_common}).


Apart from the extremely long-running queries, we observe that PostgreSQL achieves almost all performance improvements, with very few queries experiencing performance degradation. This means that specifying the join order and operators is the primary source of improvement for PostgreSQL.

\smartpara{Conclusion} Specifying join order and operators speeds up PostgreSQL, but not DuckDB. Combining the analysis of the previous two subsections~\ref{eval:overall} and ~\ref{eval:updating_stats}, join reordering is the primary source of improvement for AQP-PostgreSQL, but AQP-DuckDB mainly benefits from updating the cardinality.

\subsection{Performance Effect of the Plan Splitter}\label{eval:splitter}


In this section, we aim to answer \textbf{RQ4} by analyzing the effect of the \textbf{plan splitter} module, where the only difference is splitting the plan or not. As we described in Section~\ref{impl:pg_merge_sub_plan} and Section~\ref{impl:duckdb_merge_sub_plan}, we introduce the sub-plan combiner module after collecting the physical sub-plans from each iteration and combining them with the same execution order. We measure the plan-based AQP with the sub-plan combiner as the baseline (cf. Figure~\ref{fig:5_5_wo_split}) and compare it to the plan-based AQP (cf. Figure~\ref{fig:5_2_AQP}). As described in Section~\ref{eval:join_order}, we only measure the execution time, as the overhead of merging sub-plans is included in the end-to-end time.

\smartpara{Overall}
Both AQP-DuckDB and AQP-PostgreSQL incur performance overhead due to plan splitting (cf. Figure~\ref{fig:q4_fig_common}).
Comparing the performance effect among different benchmarks, AQP-PostgreSQL suffers the most serious degradation in the JOB benchmark ($0.51 \times$), but is least affected in DSB with a scale factor of 100 ($0.95 \times$). The performance loss of AQP-DuckDB ranges from $0.54 \times$ to $0.87 \times$.


\smartpara{Drilldown}
It is unclear whether the performance degradation is due to extreme queries. We address this by first analyzing the execution time distribution, but there are no apparent outliers. 
This indicates that the primary source of performance overhead of the plan splitter module is not from a few long-running queries. To understand the major cause of the overhead, we examine the query-by-query performance improvement in a violin plot (cf. Figure~\ref{fig:qbq_compare_q4_fig}) to determine the distribution of performance changes for each system and benchmark.

\smartpara{Distribution Study}
Both AQP-DuckDB and AQP-PostgreSQL suffer a performance loss (cf. Figure~\ref{fig:qbq_compare_q4_fig}) among both benchmarks. This indicates that despite having the same plan order, the merged plan is faster for all benchmarks to varying degrees. 
This is because the plan-based AQP method splits without knowledge of the pipeline breaker points. By predicting the pipeline breaker point in advance or jointly designing it with the executor, it is expected that the performance of plan-based AQP can be further improved.
From another perspective, it is observed that for both benchmarks, a few queries in AQP-PostgreSQL suffer significant performance loss, up to $10^{-4} \times$ in the JOB benchmark and $10^{-5} \times$ in DSB, and almost all queries slow down. Although AQP-DuckDB also suffers performance loss due to the plan splitting phase, it is less affected than AQP-PostgreSQL.
This reveals that plan splitting has a greater impact on PostgreSQL's performance. Considering that, as we described in Section~\ref{preliminaries:on_disk_and_in_memory}, on-disk DBMSs need to create temporary tables to store intermediate results, the I/O overhead from the plan splitter causes more severe performance loss to them.

\smartpara{Conclusion} The SOTA plan-based AQP's splitting strategy disrupts the original pipeline, resulting in additional overhead for both in-memory and on-disk DBMSs, where it is more severe for the latter due to the extra overhead of generating temporary tables for intermediate results. Future work should consider how to determine the appropriate split point or have a joint design with the executor to avoid this overhead.

\subsection{Performance Comparison of Plan- vs. Relation-based AQP}\label{eval:relation-based}

\begin{figure}[!t]
    \centering
    \includegraphics[width=\columnwidth]{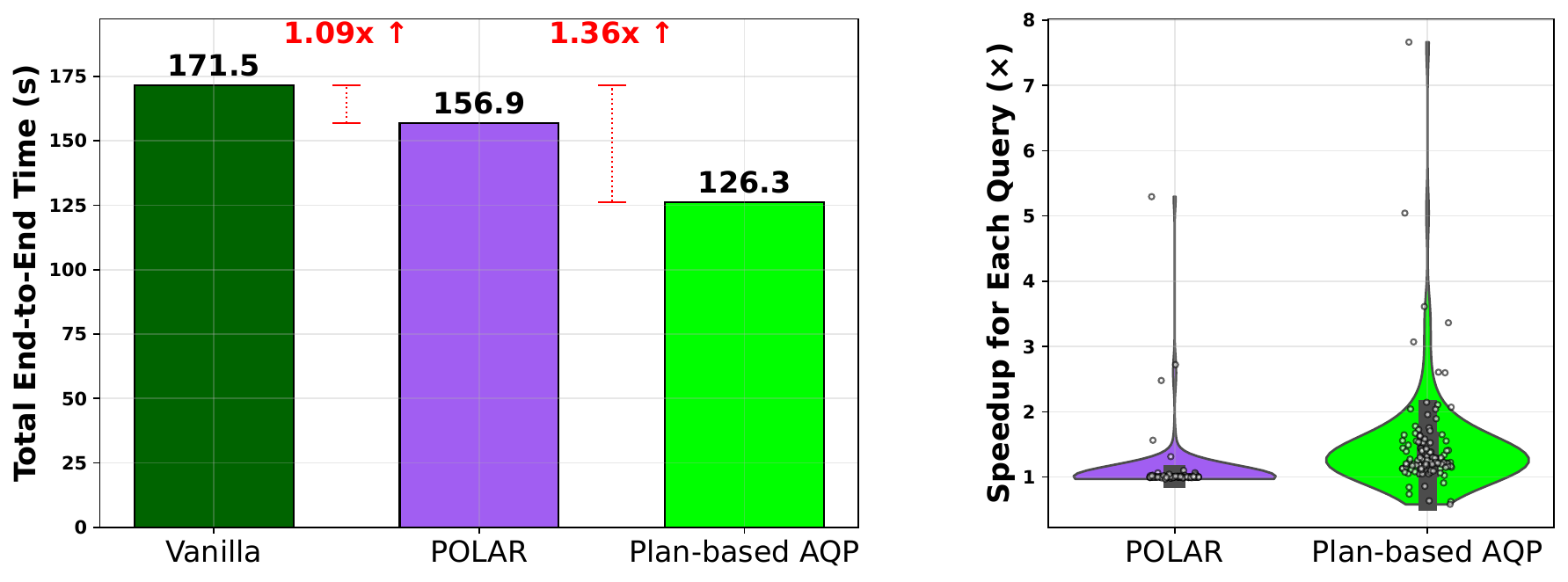}
    \caption{Performance comparison of the JOB benchmark between the Polar and the plan-based AQP strategy.}
    \label{fig:7_rq5}
    \vspace{-0.2cm}
\end{figure}

In this section, we aim to answer \textbf{RQ5} and to understand the performance improvement potential of plan-based AQP compared to relation-based AQP. We consider vanilla DuckDB (0.6.1) as the baseline. For the variants, we use Polar and AQP-DuckDB (0.6.1) for the JOB benchmark.

\smartpara{Overall}
Figure~\ref{fig:7_rq5} shows the total end-to-end time comparison results.
We observe that Polar, as a representation of the relation-based AQP, achieves $1.09\times$ speedup for the JOB. While the plan-based AQP method gains more speedup, which is $1.36\times$. However, it is not clear if the speedup is from a few extreme queries or if it is a general performance improvement.

\smartpara{Drilldown}
To address this, we analyze the performance distribution of the two systems by using violin plots (cf. Figure~\ref{fig:7_rq5}). Each chart shows the performance improvement comparing the Polar (or plan-based AQP) to the vanilla DuckDB. A value greater than 1 means performance speedup, and vice versa. 

\smartpara{Distribution Study} 
By analyzing the end-to-end time distribution, we observe that long-running queries still have more impact on the performance for both AQP methods. Compared to Polar, the plan-based AQP improves the performance of more queries, achieving up to $8\times$ speedup.

\smartpara{Conclusion}
Both relation-based and plan-based AQP improve the performance in the JOB benchmark. However, plan-based AQP achieves a much faster speedup due to the complexity of the JOB queries, which include up to 28 joins.

\section{Conclusion}\label{section:conclusion}

In this paper, we revisit the core modules of plan-based AQP methods on top of on-disk and in-memory DBMSs, analyzing how they affect performance on different benchmarks. During this process, we reproduce the SOTA with an on-disk DBMS and extend it for isolating AQP-related modules in Section~\ref{design:postgres}, including the \textbf{plan splitter}, the \textbf{sub-plan selector}, and the \textbf{monitor}. Furthermore, we apply the idea of the plan-based AQP to an in-memory DBMS by proposing a novel design in Section~\ref{design:duckdb} for overcoming the inherent flaws of the SOTA as described in Section~\ref{preliminaries:query_split}. Also, the isolation of AQP-related components is supported through our novel modular design (Section~\ref{design:components}). 

We demonstrate that, although the plan-based AQP achieves speedups for both on-disk and in-memory DBMSs (as shown in Section~\ref{eval:overall}), the primary source of this improvement differs. Plan-based AQP with the on-disk DBMS primarily benefits from a heuristic join reordering rule (as shown in Section~\ref{eval:join_order}). It can get even better performance without the AQP features (e.g., updating the "actual" cardinality from each sub-plan's execution, as shown in Section~\ref{eval:updating_stats}, and splitting the plan, as shown in Section~\ref{eval:splitter}). Still, combining the heuristic optimization rules with the AQP strategy suggests that we can go beyond the traditional plan-based AQP. Rather than locally adapting, it operates on a more global level by selecting the most suitable sub-plan (including the initial one) after each split. Based on this insight, we find that the plan-based AQP on the in-memory DBMS primarily gains speedup from updating the refined cardinality (as shown in Section~\ref{eval:updating_stats}), while avoiding the overhead of creating temporary tables, which is a common issue with on-disk DBMSs. 

Furthermore, we observe that the plan-based AQP method incurs a performance overhead due to splitting the plan within the pipeline, where on-disk DBMSs suffer more performance loss (as discussed in Section~\ref{eval:splitter}). Compared to the SOTA relation-based AQP, plan-based AQP achieves a better performance improvement, especially for the multi-join queries. We provide more details in supplementary technical report~\cite{supplementary_tech_report}.

Therefore, several potential directions for future work exist, including the development of more intelligent algorithms for selecting the split point and co-designing with the executor.

\balance

\bibliographystyle{IEEEtran}
\bibliography{my_bib}

\end{document}